\documentclass[10pt,a4paper]{article} 
\usepackage{Rgd21pap}
\begin{document}
\setcounter{page}{1}
%
%
\title{
Angular Momentum Conservative Algorithm\\
of Collisional Process in DSMC Method
}


\author{
P.A. Skovorodko\\
Institute of Thermophysics SB RAS, Novosibirsk, Russia\\
}

\maketitle 


\section{Introduction}

The traditional algorithm of collisional process in DSMC method,
used for calculation of velocities of molecules after collision,
is based on the conservation laws for linear momentum and total 
energy ~\cite{p5217r1}. This algorithm does not provide conservation
of angular momentum with respect to some axis, but for non-rotating
flow this circumstance is insignificant, since the mean value of 
angular momentum of such kind of flow is equal to zero.  \smallskip

The situation changes for axisymmetric flow with rotation. In this case 
the traditional procedure leads to some internal source or sink of 
angular momentum that may distort the flowfield. 
\smallskip

In the paper the discussed effect is recognized and investigated. The 
angular momentum conservative (AMC) algorithm of collisional process is 
proposed. The draft version of the paper is published in ~\cite{p5217r5}.

\section{Problem Formulation}

The investigations were made for one-dimensional axisymmetric flow of 
monoatomic gas in the tube with specular wall. The evolution of 
initially 
swirling flow during the time was studied. The gas in the tube with 
uniform initial density $n_0$ and temperature $T_0$ and equilibrium 
distribution function was assumed to start rotating as a solid body, the 
initial tangential velocity on the wall being equal to $v_\varphi$.
For given molecular model the problem contains two governing parameters:
the Knudsen number $Kn$, defined by the ratio of the mean free path 
$l_0$ in the gas at $t=0$ to the radius of the tube $r_t$
$$Kn\,=\,\frac{l_0}{r_t}$$

and the speed ratio 

$$S_\varphi\,=\,v_\varphi/\sqrt{2\,R\,T_0}$$ 

 The simulations were made on uniform grid with $10^4$ simulated
molecules. Standard NTC DSMC procedure ~\cite{p5217r1} was applied.  
\smallskip

The initial values of total mass $M_0$, angular
momentum $I_0$ and energy $E_0$ of the gas may be defined by 
the relations

\begin{equation}\label{p5217l1}
M_0\,=\,m\,N_0\,=\,m\,n_0\,\pi\,r^2_t
\end{equation}

\begin{equation}\label{p5217l2}
I_0\,=\,m\,\sum_{i=1}^{N_0}\,v_i\,r_i
\end{equation}

\begin{equation}\label{p5217l3}
E_0\,=\,m\,\sum_{i=1}^{N_0}\,\frac{\mbox{\boldmath$V$}_i^2}{2}
\end{equation}

where $N_0$ is the total number of simulated molecules, $m$ - the
mass of molecule, $\mbox{\boldmath$V$}$ - the velocity vector,
$v$ - the tangential component of this vector. By the same relations
may be determined the total energy $E$ and total 
angular momentum $I$ of the gas during the time.\smallskip

For any values of $Kn$ and $S_\varphi$
there should be no temporal dependence of $E$ and $I$.
The computations show that this condition is satisfied only 
for $E$.\smallskip

\begin{figure}[!ht]
\epsfxsize=\columnwidth \epsfbox{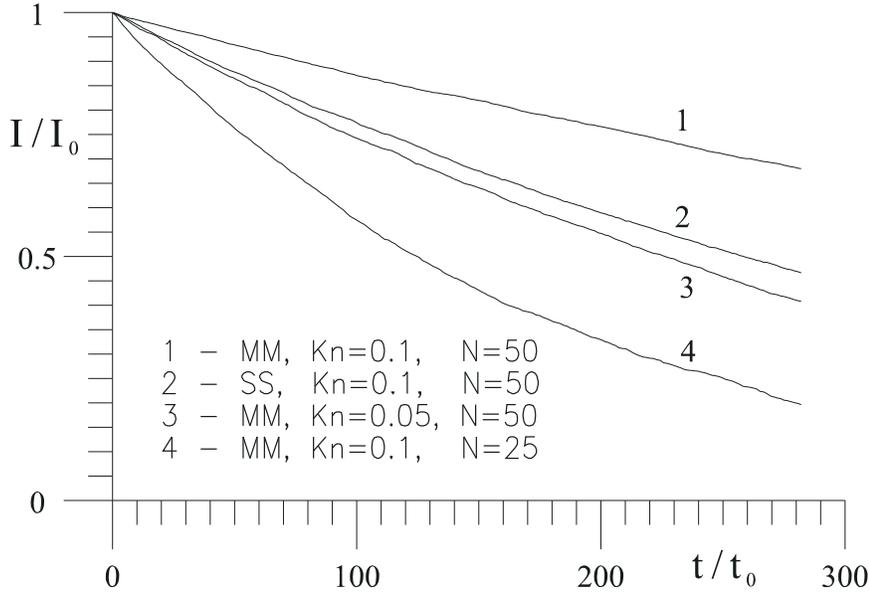}
\caption{Temporal dependence of total angular momentum
} \label{P5217f1} \end{figure}

Fig.\ref{P5217f1} illustrates the dependencies of $I/I_0$ on the time
for some of considered variants for $S_\varphi\,=\,1$. The time is normalized
by the value of $t_0\,=\,2\,\pi\,r_t/v_\varphi$, so $t/t_0$ represents the
number of revolutions of solid body with $v(r_t)=v_\varphi$.

As can be seen from Fig.\ref{P5217f1}, for all of the variants the
decreasing of $I/I_0$ with increasing of $t/t_0$ is observed. For given
$Kn$ the effect is more pronounced for solid sphere molecules $(SS)$ in 
comparison with $VSS$ molecular model for Maxwell molecules $(MM)$.
The effect depends on the Knudsen number: the less $Kn$ the more the 
effect. But the highest sensitivity of the effect is observed to the 
grid size: changing the number $N$ of cells from $50$ to $25$ causes 
the same effect as the decrease of the Knudsen number approximately 
by $4$ times.\smallskip

Due to the dependence of $I(t)$ it is difficult to obtain the steady 
solution of the problem. The dissipation of rotational contribution of
energy into the heat takes place during the time. The only steady
solution that may be obtained with traditional collisional procedure
is terminal one, when at $t\,\rightarrow\,\infty$ the gas will stop
rotating, having uniform density $n_0$ and getting warm to the temperature 
$$T\,=\,2\,E_0/\,3\,R\,M_0$$

\section{AMC Algorithm}

The reason of temporal dependence of total angular momentum is quite 
clear: the colliding molecules have different radial coordinates 
$r_1, r_2$, that leads to the difference in angular momentum 
$$m_1\,v_1\,r_1\,+\,m_2\,v_2\,r_2$$
of these molecules before and after
collision ($v$ denotes the tangential component of velocity of the
molecule). To provide the conservation of angular momentum the algorithm
of collisional process should be changed. One possible variant of AMC 
algorithm is as follows.\smallskip

Let us denote by $\mbox{\boldmath$u$}$ the vector containing axial
and radial components of the velocity of molecule. The post-collision
velocities calculated by traditional procedure will be denoted by 
symbol $*$, while the symbol $'$ will be used for these velocities
in the proposed algorithm.\smallskip

The algorithm is based on conservation law of angular momentum 
$m\,v\,r$ for tangential component of velocity instead of linear
momentum $m\,v$. That is why the post-collision tangential 
velocities are calculated by usual relations ~\cite{p5217r1},
with the value of $m_i$ being replaced by the product $m_i\,r_i$.
The obtained velocities $v'_1, v'_2$ provide precise conservation 
of angular momentum, but violate the energy conservation.
To provide energy conservation 
the correction of velocities $\mbox{\boldmath$u$}^*_1$
and $\mbox{\boldmath$u$}^*_2$ is needed.
To do this correction the analysis of energy defects 
$$dE_i\,=\,m_i\,(v'^2_i-v^{*2}_i)$$
should be made. Three possible relations 
between energy defects $dE_i$ and velocities 
$\mbox{\boldmath$u$}^*_i$ are as follows.\smallskip

1. If 
$$m_1\,u^{*2}_1\,>\,dE_1$$
and
$$m_2\,u^{*2}_2\,>\,dE_2$$
both components of vectors $\mbox{\boldmath$u$}^*_i$ should be
corrected by the factors $c_i$
$$c_i\,=\,\sqrt{1\,-\,dE_i/m_i\,u^{*2}_i}$$
\smallskip

2. If 
$$m_1\,u^{*2}_1\,+\,m_2\,u^{*2}_2\,>\,dE_1\,+\,dE_2$$ 
the correction factor $c$ for components of both velocities
$\mbox{\boldmath$u$}^*_i$ should be the same and equal to 
$$c\,=\,\sqrt{1\,-\,(dE_1\,+\,dE_2)/(m_1\,u^{*2}_1\,+\,m_2\,u^{*2}_2)}$$
\smallskip

3. If none of the above conditions is satisfied, the collision is 
considered to be "bad" and is not performed, i. e. the molecules 
conserve their pre-collision velocities.\smallskip

The computations show that the relative number of collisions of types 1 
and 2 is about $99\%$ and $1\%$, respectively. The relative number of 
"bad" collisions never exceeds $10^{-4}$, so the effect caused by the
neglecting of these collisions is small. 

\section{Steady Solution}

The described algorithm provides precise conservation of angular 
momentum and energy. All the curves, presented in Fig.\ref{P5217f1} 
transform into straight line $I/I_0\,=\,1$, if this algorithm is applied.  
\smallskip

AMC algorithm enables to obtain the steady solution of the
considered problem. This solution proves to be the same, as the 
prediction of Navier-Stokes theory and has the following features.
\smallskip

1. The flow is isothermal

\begin{equation}\label{p5217l4}
T\,=\,const  
\end{equation}

2. The gas rotates as a solid body 
\begin{equation}\label{p5217l5}
v\,=\,v_w\,r/r_t  
\end{equation}

3. The radial distribution of density is described by the relation
\begin{equation}\label{p5217l6}
n(r)\,=\,n(0)\,\exp{(v^2/2\,R\,T)}
\end{equation}

4. The solution is completely determined by the values of $M_0$, $I_0$, 
$E_0$ and does not depend on the way of initial swirling.
\smallskip

5. The solution does not depend either on the molecular model or
on the Knudsen number.\smallskip

These features of the considered flow are important for 
clarifying the nature of the Ranque effect 
~\cite{p5217r2, p5217r3}.\smallskip

Three unknown values $v_w, T, \,n(0)$ determining the radial distribution
of parameters of steady solution may be defined based on the values of
$M_0$, $I_0$ and $E_0$ from the relations
\begin{equation}\label{p5217l7}
\pi\,r^2_t\,m\,n(0)\,\frac{(\exp{S^2_w}-1)}{S^2_w}\,=M_0
\end{equation}
\smallskip
\begin{equation}\label{p5217l8}
M_0\,v_w\,r_t\,\frac{\exp{S^2_w}\,(S^2_w-1)+1}{S^2_w\,
(\exp{S^2_w}-1)}\,=\,I_0
\end{equation}
\smallskip
\begin{equation}\label{p5217l9}
\frac{3}{2}\,R\,T\,M_0\,+\,\frac{v_w\,I_0}{2\,r_t}\,=\,E_0
\end{equation}
where speed ratio $S_w$ of the flow at $r\,=\,r_t$ is defined by
\begin{equation}\label{p5217l10}
S_w\,=\,v_w/\sqrt{2\,R\,T} 
\end{equation}

The left hand sides of the relations \ref{p5217l7}\,-\,\ref{p5217l9}
representing the total mass, angular momentum and energy of the flow
may be obtained based on the relations \ref{p5217l4}\,-\,\ref{p5217l6}
by simple integration.
\smallskip

\begin{figure}[!ht]
\epsfxsize=\columnwidth \epsfbox{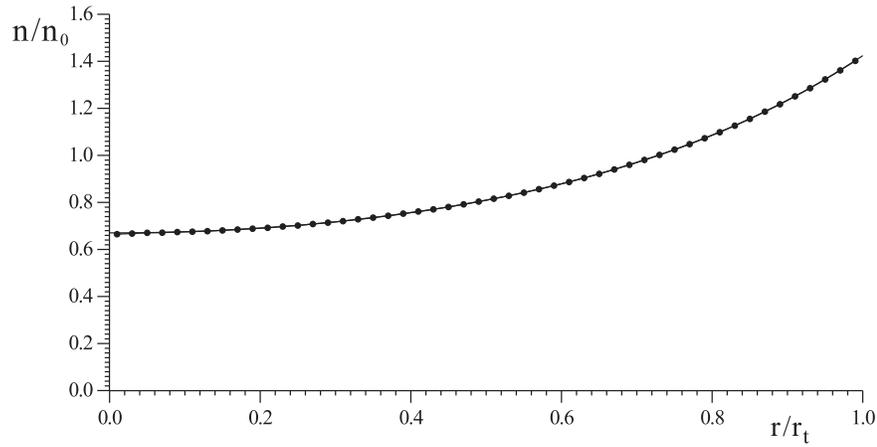}
\caption{Radial distribution of density
} \label{P5217f2} \end{figure}

\begin{figure}[!ht]
\epsfxsize=\columnwidth \epsfbox{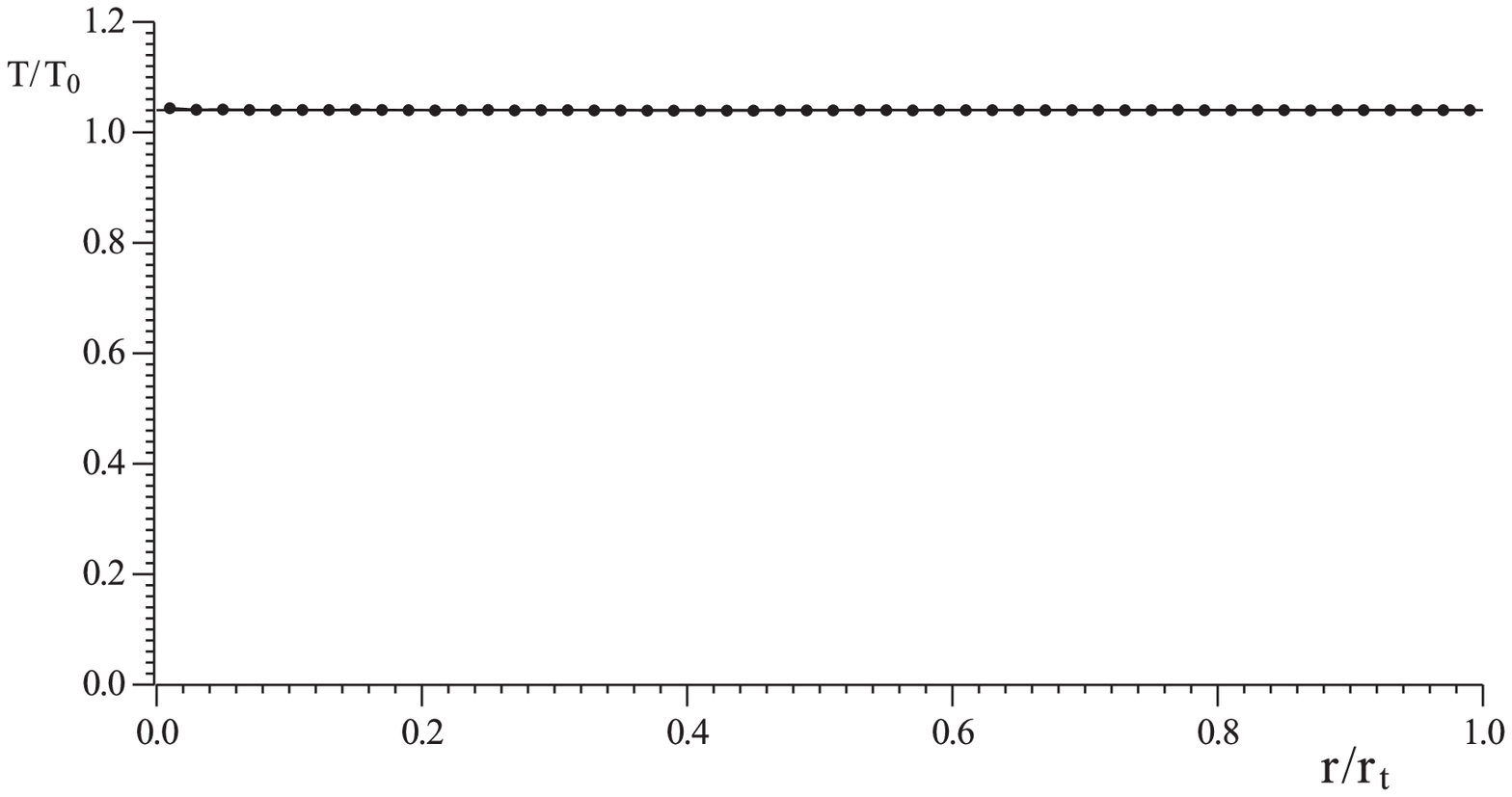}
\caption{Radial distribution of temperature
} \label{P5217f4} \end{figure}

The comparison between numerical and analytical radial distribution
of parameters of steady flow for $VSS$ molecules, $Kn\,=\,0.1$, 
$S_\varphi\,=\,1$, $N\,=\,50$ is made on Figs. 
\ref{P5217f2}\,-\,\ref{P5217f4} for density, tangential velocity and
temperature, respectively. The numerical results are shown by solid
circles while solid lines represent the results of analytical solution 
\ref{p5217l7}\,-\,\ref{p5217l9}. As can be seen from these Figures, 
numerical and analytical results are in excellent agreement. For
$Kn\,=\,10$ the steady solution was found to be the same as for
$Kn\,=\,0.1$.\smallskip

It should be noted that the steady flow in the considered problem
is characterized by locally Maxwellian distribution function for
any molecular models and $Kn\,<\,\infty$.\smallskip

\begin{figure}[!ht]
\epsfxsize=\columnwidth \epsfbox{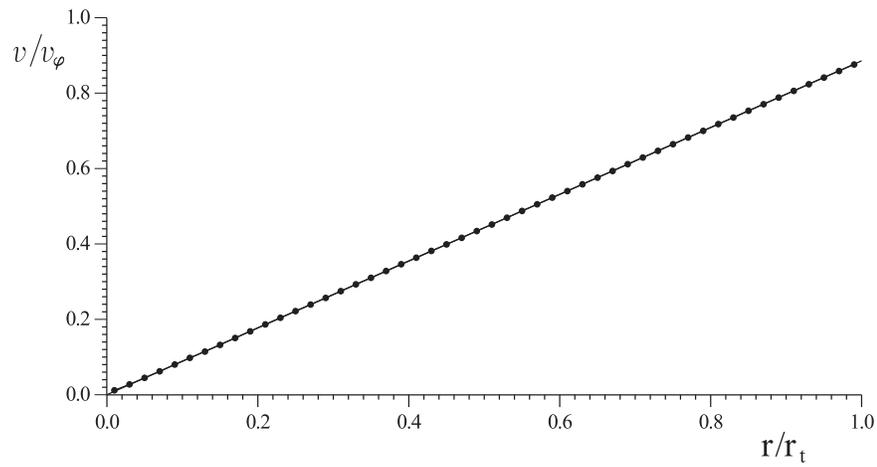}
\caption{Radial distribution of tangential velocity
} \label{P5217f3} \end{figure}

AMC algorithm of collisional process may be applied for 
simulation of non-rotating flows too. The experience of such application 
to the flow in gasdynamic window ~\cite{p5217r4} shows that the results 
obtained with proposed and traditional algorithms are the same within the 
expected statistical scatter.\smallskip

\begin{figure}[!ht]
\epsfxsize=\columnwidth \epsfbox{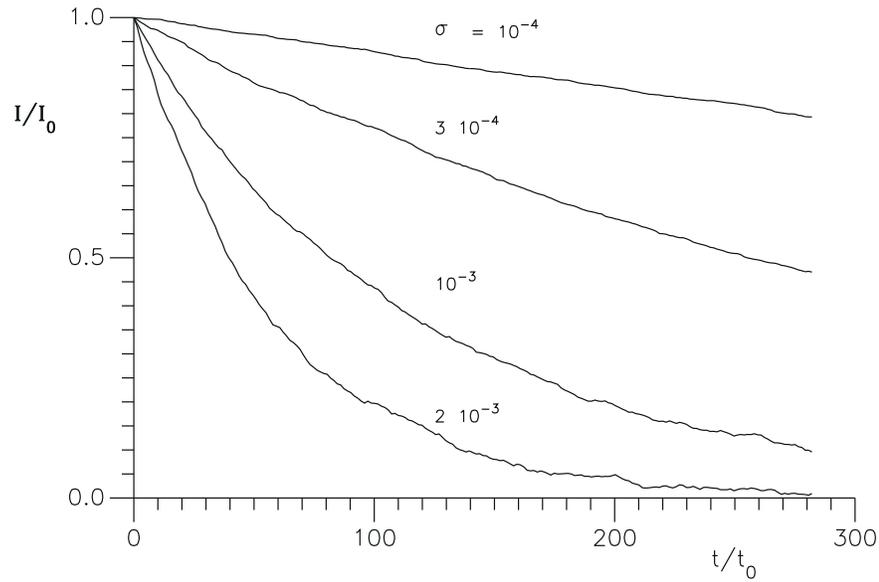}
\caption{Temporal dependence of total angular momentum
} \label{P5217f5} \end{figure}

To stay the effect of non-conservative nature of traditional 
collisional procedure in the real conditions, some test computations
of considered flow were made by AMC algorithm for adiabatic
tube wall with diffuse-specular scattering of molecules for different
values of accommodation coefficient $\sigma$. Fig.\ref{P5217f5}
illustrates the dependencies of $I/I_0$ on the time
for $VSS$ molecules, $Kn\,=\,0.1$, $S_\varphi\,=\,1$, $N\,=\,50$ and
$4$ values of $\sigma$ in the range $\sigma\,=\,2\cdot10^{-3}\,-\,10^{-4}$.
The comparison of these dependencies with those presented in 
Fig.\ref{P5217f1} reveals that the traditional collisional procedure
causes the same change in angular momentum as the diffuse-specular wall
with the accommodation coefficient $\sigma$ in the range 
$\sigma\,=\,10^{-3}\,-\,10^{-4}$. Therefore, for flows with real
surface  $(\sigma\,\sim\,1)$ in the flowfield the studied effect 
is small, its relative value is about $10^{-3}\,-\,10^{-4}$.

\section{Conclusion}

The proposed algorithm of collisional process provides precise 
conservation of angular momentum and energy and may be used for
direct simulation of flows with rotation.\smallskip

It should be noted, however, that the improvement of the results,
obtained by this algorithm may be important only for the problems
without external source or sink of angular momentum. 
If solid surface with diffuse reflection is present in the flowfield, 
the changes in angular momentum, caused by this surface will be
much greater than those, caused by non-conservative nature of traditional
algorithm of collisional process.  For such kind of flows there is 
no need to employ AMC algorithm, especially taking into account
that it takes $10\%\,-\,15\%$ more CPU time.\smallskip

Nevertheless, for some problems the application of AMC algorithm is 
advisable. An example of such kind of problem is the Ranque effect
~\cite{p5217r2}, which nature is determined by conservation of angular 
momentum of the gas flow in the vortex tube ~\cite{p5217r3}.

\end{document}